# The magnetic field of pulsars and the gravito-magnetic theory

Jacob Biemond*

*Vrije Universiteit, Amsterdam, Section: Nuclear magnetic resonance, 1971-1975*

## Abstract

Many authors have considered a gravitational origin of the magnetic field of celestial bodies. Especially, the so-called Wilson-Blackett formula has been investigated, both theoretically and observationally. It appeared possible to deduce this formula from general relativity, e. g., by application of a special interpretation of gravito-magnetic theory. More consequences of the latter theory for pulsars will be considered in this work.

As an example, the standard quadrupolar charge density for pulsars can be deduced from the gravito-magnetic theory. Moreover, a new magnetic dipole moment from electromagnetic origin is found, generated in the basic magnetic field from gravito-magnetic origin.

In general, for thirteen accreting, slowly rotating, binary pulsars the agreement between the observed magnetic field and the gravito-magnetic prediction is better than between the observed value and the value from the standard magnetic dipole radiation model. At present, an analogous comparison for five isolated pulsars appears to be difficult.

For a sample of 100 pulsars the averaged *(gravito-)*magnetic field, extracted from the magnetic dipole spin-down model, is in fair agreement with the gravito-magnetic prediction. Unfortunately, the *(gravito-)*magnetic field has not yet *directly* been measured.

Finally, it is found that the first and second order braking indices only depend on the magnetic field from electromagnetic origin.

## 1. Introduction

Since 1891 many authors have discussed a gravitational origin of the magnetic field of rotating celestial bodies. Especially, the so-called Wilson-Blackett formula was considered [1–8]

$$\mathbf{M} = -1/2 \, \beta \, c^{-1} \, G^{1/2} \, \mathbf{S}. \tag{1}$$

Here $\mathbf{M}$ is the magnetic dipole moment of the massive body with angular momentum $\mathbf{S}$, $c$ is the velocity of light in vacuum, $G$ is the gravitational constant and $\beta$ is a dimensionless constant of order unity. Attempts to derive equation (1) from a more general theory have been given by several authors [8–14]. Equation (1) has often been considered as a consequence of general relativity [12–14], for example, in a special version of the gravito-magnetic theory [8]. In the latter approach the so-called "magnetic-type" gravitational field is identified as a *common* magnetic field, resulting into the magnetic dipole moment $\mathbf{M}(gm)$ of (1).

Experiments on rotating masses in the laboratory in order to test (1) have been performed by Blackett [15] and others [16–18]. Available observations and theoretical considerations with respect to the relation (1) and other explanations of the origin of the magnetic field of celestial bodies have been reviewed by Biemond [8]. Following Woodward [19], the validity of the Wilson-Blackett (or Schuster-Blackett) formula for pulsars has again been investigated in this work, but now from the gravito-magnetic point of view.

---

*Postal address: Sansovinostraat 28, 5624 JX Eindhoven, The Netherlands.
Website: http://www.gewis.nl/~pieterb/gravi/ , e-mail: gravi@gewis.nl

The angular momentum **S** for a sphere of radius $R$ can be calculated from the relation

$$\mathbf{S} = I\,\mathbf{\Omega}, \text{ or } S = I\,\Omega = 2/5\,f\,m\,R^2\,\Omega, \qquad (2)$$

where $m$ is the mass of the sphere, $\Omega = 2\pi\,P^{-1}$ its angular velocity ($P$ is the rotational period), $I$ its moment of inertia and $f$ is a dimensionless factor depending on the homogeneity of the mass density in the sphere (for a homogeneous mass distribution $f = 1$). The value of the magnetic dipole moment **M** may be calculated from

$$\mathbf{M} = \tfrac{1}{2}\,r^3\,\mathbf{B}_p, \text{ or } M = \tfrac{1}{2}\,r^3\,B_p. \qquad (3)$$

Here $\mathbf{B}_p$ is the magnetic induction field at the pole of the sphere and $r$ is the distance from the centre of the sphere to the field point where $\mathbf{B}_p$ is measured. The magnetic moment $M$ has been derived for $r > R$, but in this work values of $B_p$ measured at $r \approx R$ have been introduced into (3).

Combination of (1)–(3) yields the following gravito-magnetic prediction for $\mathbf{B}_p$

$$\mathbf{B}_p(gm) = -\beta\,c^{-1}\,G^{1/2}\,I\,R^{-3}\,\mathbf{\Omega}, \text{ or } B_p(gm) = -5.414 \times 10^{13}\,P^{-1}, \quad (\beta = +1) \qquad (4)$$

where the minus sign reflects that the vectors $\mathbf{B}_p(gm)$ and $\mathbf{\Omega}$ possess opposite directions for $\beta = +1$. Neither the sign nor the value of $\beta$ follow from the gravito-magnetic theory. Therefore, absolute values of $B_p(gm)$ are given in this work. The representative values $I = 10^{45}$ g.cm$^2$ and $R \approx r = 10$ km have been inserted in (4) and in other formulas in this work. It is noticed that (4) imposes no restriction on the maximum of $B_p(gm)$, i.e., $B_p(gm)$ may be larger than the critical field strength $m_e^2\,c^3/e\,\hbar = 4.414 \times 10^{13}$ Gauss, at which electron-positron pair creation processes may become probable.

As pointed out earlier [8, pp. 12, 14, 19–20 and 49], moving electric charge in the magnetic field from gravito-magnetic origin may cause an additional magnetic field from electromagnetic origin. The latter field may partly or completely compensate the magnetic field from gravito-magnetic origin. It is noticed that the magnetic field generated by rotating neutral mass is generally much smaller than the magnetic field generated by moving charge. For example, for the electron one may compare the following magnetic moment to angular momentum ratios: $(M/S)_{gravito\text{-}magnetic}$ and $(M/S)_{electromagnetic}$. Calculation shows: $(M/S)_{gravito\text{-}magnetic}/(M/S)_{electromagnetic} = G^{1/2}\,m_e\,e^{-1} = 4.899 \times 10^{-22}$. Therefore, as a rule, proposed magnetic fields from gravito-magnetic origin are extremely small and difficult to isolate from fields of electro-magnetic origin.

When a magnetic induction field $\mathbf{B}(gm)$ due to gravito-magnetism is present, the total magnetic induction field $\mathbf{B}(tot)$ is given by

$$\mathbf{B}(tot) = \mathbf{B}(gm) + \mathbf{B}(em), \qquad (5)$$

where $\mathbf{B}(em)$ is the magnetic induction field due to electromagnetism. It may be helpful to define the dimensionless quantities $\beta_{eff}$ and $\beta'$ by the relations

$$\mathbf{B}(tot) \equiv \beta_{eff}\,\mathbf{B}(gm), \text{ and } \mathbf{B}(em) \equiv \beta'\,\mathbf{B}(gm). \qquad (6)$$

Since the signs of $\beta_{eff}$ and $\beta'$ are unknown, absolute values of these quantities have been given in all tables below. Combination of (5) and (6) yields

$$\mathbf{B}(em) = (\beta_{eff} - 1)\,\mathbf{B}(gm), \text{ and } \beta_{eff} = 1 + \beta'. \qquad (7)$$

When the total field $\mathbf{B}(tot)$ is only due to gravito-magnetic origin, $\mathbf{B}(em) = 0$ and $\beta_{eff}$ and $\beta'$ reduce to $\beta_{eff} = 1$ and $\beta' = 0$, respectively.

Since charges may move in different ways in pulsars and other rotating bodies, one can hardly expect that $β_{eff}$ is a constant. Indeed, different results for $β_{eff}$ have been found for about fourteen rotating bodies: metallic cylinders in the laboratory, moons, planets, stars and the Galaxy. However, for this series a mean value of $β_{eff} = 0.076$ has been calculated [8]. Although the agreement with the gravito-magnetic prediction $β_{eff} = 1$ is only approximate, the correct order of magnitude of $β_{eff}$ for so many, *strongly different*, rotating bodies is amazing (the values of the parameters $m$, $R$, $Ω$ and $B_p$ occurring in (4) (see (1) and (6a)) can *differ by many decades*). Such a result may reflect the validity of the gravito-magnetic hypothesis.

In general, many properties of pulsars are determined by the emission model used. Therefore, we will first pay attention to some proposed emission models in section 2. In sections 3 and 6 the influence of the magnetic field on first and second order braking indices is considered, both theoretically and observationally, respectively. Apart from the Wilson-Blackett formula, other consequences of the gravito-magnetic theory for pulsars are deduced in section 4. In section 5 observational data for different kinds of pulsars are summarized and compared with theoretical predictions. A summary of the conclusions is given in section 7.

## 2. Emission models of pulsars

The emission model of pulsars is an important factor in the prediction of the magnetic field of pulsars. Unfortunately, no generally accepted model is available. Therefore, we will discuss some alternative formulas for the emission of pulsars.

The rotational energy $E = ½IΩ^2$ of the pulsar may change by magnetic dipole radiation, accretion or other mechanisms according to

$$\dot{E} = I\,Ω\,\dot{Ω} = -4π^2\,I\,P^{-3}\,\dot{P}, \tag{8}$$

where $\dot{E}$, $\dot{Ω}$ and $\dot{P}$ are the time-derivatives of $E$, $Ω$ and $P$, respectively (It has been assumed that the moment of inertia $I$ is no function of time). The contribution of the *(electro)*magnetic dipole radiation in vacuum to $\dot{E}$ of (8) is given by (see, e. g., [20, pp. 188–189; 21, pp. 176–177])

$$\dot{E} = -⅔\,c^{-3}\sin^2θ\,M(\text{em})^2\,Ω^4 = -(32π^4/3)\,c^{-3}\sin^2θ\,M(\text{em})^2\,P^{-4}, \tag{9}$$

where $M(\text{em})$ is the *(electro)*magnetic dipole moment of the pulsar and $θ$ is the angle between **M**(em) and **S**. Note that the *(gravito-)*magnetic dipole moment **M**(gm) does not occur in (9), since *(gravito-)*magnetic dipole radiation does not exist. Gravito-magnetic quadrupole radiation (for two point masses its energy formula coincides with the familiar expression for the gravitational quadrupole radiation (see, [8, ch. 3; 20]) may be present, however, but its influence is not considered in this work.

When $\dot{E}$ in (8b) is taken equal to $\dot{E}$ in (9b), it follows that $\dot{P}$ is *positive*. So, magnetic dipole radiation leads to *spin-down*. Combination of (8b) and (9b) further yields for $M(\text{em})$

$$M(\text{em}) = \left(\frac{3\,c^3\,I}{8π^2\sin^2θ}\right)^{1/2} (P\,\dot{P})^{1/2}. \tag{10}$$

Choosing $θ = 90°$ in (10), the following magnetic induction field at the pole of the pulsar can be calculated from (3b)

$$B_p(\text{sd}) = \left(\frac{3\,c^3\,I}{2π^2\,R^6}\right)^{1/2} (P\,\dot{P})^{1/2} = 6.399 \times 10^{19}\,(P\,\dot{P})^{1/2}. \tag{11}$$

It is noticed that, as a rule, $\theta < 90°$, so that $B_p(sd)$ in (11) must be considered as the lower limit.

Assuming deformation of magnetic field lines, or outflow of high-energy charged particles, approximations to $\dot{E}$ may be obtained by taking the product of the energy density of the equatorial *(electro)*magnetic field at the light cylinder $B_L(em)$, the effective area of the light cylinder and the velocity of light (see, e. g., [21, pp. 188–189])

$$\dot{E} = -\frac{B_L(em)^2}{8\pi} 4\pi R_L^2 c, \tag{12}$$

where $R_L = c/\Omega$ is the radius of the light cylinder. $B_L(em)$ is taken equal to

$$B_L(em) = \tfrac{1}{2} B_p(em) (R/R_L)^p, \tag{13}$$

where $B_p(em)$ is the *(electro)*magnetic field at the pole of the pulsar. Using (3b), combination of (12) and (13) leads to

$$\dot{E} = -\tfrac{1}{8} c^{3-2p} R^{2p} B_p(em)^2 \Omega^{2p-2} = -\tfrac{1}{2} c^{3-2p} R^{2p-6} M(em)^2 \Omega^{2p-2}. \tag{14}$$

For example, $p = 3$ yields

$$\dot{E} = -\tfrac{1}{2} c^{-3} M(em)^2 \Omega^4. \tag{15}$$

Although the derivations of (9a) and (15) start from different assumptions, both equations nearly coincide, apart from a different coefficient and a factor $\sin^2\theta$. Since the derivation of (9a) is more rigorous than that of (15), the former expression is usually used as the standard formula for the determination of the *(electro)*magnetic field of pulsars.

When $p = 2$, equation (14) yields for $\dot{E}$

$$\dot{E} = -\tfrac{1}{2} c^{-1} R^{-2} M(em)^2 \Omega^2 = -\tfrac{1}{8} c^{-1} R^4 B_p(em)^2 \Omega^2. \tag{16}$$

Combination of (8b) and (16b) then leads to the following expression for the deformed magnetic induction field, denoted by $B_p(def)$

$$B_p(def) = \left(\frac{8\,c\,I}{R^4}\right)^{1/2} \left(\frac{\dot{P}}{P}\right)^{1/2} = 1.549 \times 10^{16} \left(\frac{\dot{P}}{P}\right)^{1/2}. \tag{17}$$

The calculated values of $B_p(gm)$ of (4), $B_p(sd)$ of (11) and $B_p(def)$ of (17) will be compared with each other and with available observational data below.

## 3. Magnetic field dependence of braking indices

In order to calculate so-called braking indices, equations (8a) and (14a) are combined to

$$\dot{\Omega} = -\tfrac{1}{8} c^{3-2p} I^{-1} R^{2p} B_p(em)^2 \Omega^{2p-3}. \tag{18}$$

Alternatively, (18) may be generalized to

$$\dot{\Omega} = -k_p\, c^{3-2p} I^{-1} R^{2p} B_p(em)^2 \Omega^{2p-3}, \tag{19}$$

where $k_p$ is a constant factor. In the sequel of this work it will be assumed that the quantity $p$ is no function of time. Direct combination of (3b), (8a), (9a) and (19) shows that $k_3 = 1/6 \sin^2\theta$ for $p = 3$.

Instead of using (18) and (19), Kaspi *et al.* [22] and Johnston and Galloway [23] followed Manchester and Taylor [21] and considered the generalized relation

$$\dot{\Omega} = -K\Omega^{n_0} = -k B_p(\text{em})^2 \Omega^{n_0}, \tag{20}$$

where the quantities $k$ and $n_0$ are assumed not to depend on time.

From (18)–(20) the quantity $\ddot{\Omega}$ and the so-called first order braking index $n$ can be calculated by *differentiation* (compare with the *integration* method discussed in section 6). From (18) and (19) one obtains

$$n \equiv \frac{\Omega \ddot{\Omega}}{\dot{\Omega}^2} = 2p - 3 + 2\frac{\dot{B}_p(\text{em})\,\Omega}{B_p(\text{em})\,\dot{\Omega}}, \tag{21}$$

where the first order braking index $n$ is defined in terms of the observable quantities $\Omega$, $\dot{\Omega}$ and $\ddot{\Omega}$. In deriving (21) it has been assumed that none of the following parameters *depend on time*: the moment of inertia $I$, the radius $R$, the mass $M$, or the angle $\theta$ of the pulsar. For convenience sake, these assumptions will be also used in the sequel of this paper. Note that for a fixed value of $p$ the braking index $n$ will become smaller for $B_p(\text{em}) < 0$ and $\dot{B}_p(\text{em}) < 0$, since $\Omega > 0$ and usually $\dot{\Omega} < 0$.

From (20b) follows, analogously to (21)

$$n \equiv \frac{\Omega \ddot{\Omega}}{\dot{\Omega}^2} = n_0 + 2\frac{\dot{B}_p(\text{em})\,\Omega}{B_p(\text{em})\,\dot{\Omega}}. \tag{22}$$

Comparison of (21) and (22) shows that $n_0 = 2p - 3$. Note that, in principle, $n$ and $n_0$ need not to be integers.

It is noticed that the braking indices $n$ of (21) and (22) neither directly depend on the gravito-magnetic field $B_p(\text{gm})$ nor on $\dot{B}_p(\text{gm})$. *Therefore, the validity of the gravito-magnetic hypothesis cannot been tested by considering the braking index n.*

In the evaluation of (18)–(20) it is often assumed that $B_p(\text{em})$ is *no function of time*. In that case $n$ in (21) reduces to $n = 2p - 3$, whereas (22) simplifies to $n = n_0$.

In addition, one may identify the field $B_p(\text{em})$ in (22) with the field $B_p(\text{sd})$ from (11) that depends on the quantities $\Omega$ and $\dot{\Omega}$. Differentiation of $B_p(\text{em}) = B_p(\text{sd})$ with respect to time followed by evaluation yields for the ratio $\dot{B}_p(\text{sd})$ to $B_p(\text{sd})$

$$\frac{\dot{B}_p(\text{sd})}{B_p(\text{sd})} = \tfrac{1}{2}(n-3)\frac{\dot{\Omega}}{\Omega} = \tfrac{1}{4}(3-n)\frac{1}{\tau_c}. \tag{23}$$

Here the quantity $\tau_c$ is defined by $\tau_c \equiv -\tfrac{1}{2}\Omega/\dot{\Omega} = \tfrac{1}{2} P/\dot{P}$. When $n = 3$, it follows from (23) that $B_p(\text{sd})$ is independent of time, but for $n \neq 3$ $\dot{B}_p(\text{sd}) \neq 0$. When $n < 3$ and $B_p(\text{sd}) < 0$, $B_p(\text{sd})$ becomes more negative in time, whereas for $n > 3$ and $B_p(\text{sd}) < 0$ $B_p(\text{sd})$ becomes less negative in time.

Combination of (22) and (23a) shows that $n_0 = 3$. Therefore, the quantity $n_0$ has been called the true braking index (see, e. g., [22]). Moreover, from the relation $n_0 = 2p - 3$ follows $p = 3$ in this case. Summing up, *(electro)magnetic dipole radiation implies $n_0 = 3$ and $p = 3$*, but the value of $n$ is determined by the time dependence of $B_p(\text{sd})$ (or more precisely by the ratio $\dot{B}_p(\text{sd})/B_p(\text{sd})$: see equation (23)). Note that in favourable cases the braking index $n$ can be calculated from $n = \Omega\ddot{\Omega}/\dot{\Omega}^2$, where $n$ need not to be an integer.

When $B_p(\text{em})$ is identified with $B_p(\text{sd})$, it is appears possible to deduce an expression for the factor $\beta'$ in (6b) by combining (4), (6b) and (11)

$$\beta' = (-6\,c^5\,G^{-1}\,I^{-1}\,\Omega^{-5}\,\dot{\Omega})^{1/2}. \tag{24}$$

This result shows that $\beta'$ depends on the quantities $\Omega$ and $\dot{\Omega}$, and on the particular choice $B_p(\text{em}) = B_p(\text{sd})$.

Another possibility is to identify $B_p(\text{em})$ with $B_p(\text{def})$ from (17). Differentiation of $B_p(\text{em}) = B_p(\text{def})$ with respect to time yields

$$\frac{\dot{B}_p(\text{def})}{B_p(\text{def})} = \tfrac{1}{2}(n-1)\frac{\dot{\Omega}}{\Omega} = \tfrac{1}{4}(1-n)\frac{1}{\tau_c}. \tag{25}$$

When in this case $n = 1$, it follows from (25) that $B_p(\text{def})$ is independent of time. When $n < 1$ and $B_p(\text{def}) < 0$, $B_p(\text{def})$ becomes more negative in time, whereas for $n > 1$ and $B_p(\text{def}) < 0$ $B_p(\text{def})$ becomes less negative in time. Combination of (22) and (25a) shows that now $n_0 = 1$. From the relation $n_0 = 2p - 3$ then follows that $p = 2$. Summing up, in the case of deformation of magnetic field lines, or outflow of high-energy charged particles, $n_0$ and $p$ may have the values $n_0 = 1$ and $p = 2$. Again the observed braking index $n$ need not to be an integer.

In table 1 some results for $n$ and $n_0$ from (21) and (22) with respect to the time dependence of $B_p(\text{em})$ are summarized. See the text for further information.

**Table 1. Braking indices $n$ and $n_0$ depending on the emission model ($p$-value) and the field $B_p(\text{em})$.**

|  | $B_p(\text{em}) \neq f(t)$ | $B_p(\text{em}) = B_p(\text{sd})$ | $B_p(\text{em}) = B_p(\text{def})$ |
|---|---|---|---|
| General case $p = p$ | $n = 2p - 3$ $n = n_0$ | $n_0 = 2p - 3$ | $n_0 = 2p - 3$ |
| Magnetic dipole radiation $p = 3$ | $n = n_0 = 3$ | $n_0 = 3$ | – |
| Deformation of field lines $p = 2$ | $n = n_0 = 1$ | – | $n_0 = 1$ |

In order to get an idea of the evolution a pulsar, the true age $t$ of a pulsar is an important parameter. If $B_p(\text{em})$ is *no function of time*, integration of (18) or (19) yields the following expression for this quantity (compare with, e.g., [21, p. 111])

$$t = \frac{-\Omega}{(n-1)\dot{\Omega}}\left\{1 - \left(\frac{\Omega}{\Omega_i}\right)^{n-1}\right\} \approx \frac{-\Omega}{(n-1)\dot{\Omega}} = \frac{2\tau_c}{n-1}, \quad (n \neq 1) \tag{26}$$

where $\Omega_i$ is the initial value of $\Omega$ at time $t = 0$ and $\tau_c \equiv -\tfrac{1}{2}\Omega/\dot{\Omega} = \tfrac{1}{2}P/\dot{P}$. According to (21), $n$ then equals to $n = 2p - 3$ and neither $n$ nor $p$ need not to be integers. From (26b) it can be seen that the characteristic time $\tau_c$ may be used as an estimate for $t$. Therefore, $\tau_c$ has been added to tables 2 through 4 when $P$ and $\dot{P}$ are known.

Following Kaspi *et al.* [22], a second order braking index $m$ can be calculated from (20b). Taking the time dependence of $B_p(\text{em})$ into account, one finds

$$m \equiv \frac{\Omega^2 \ddot{\Omega}}{\dot{\Omega}^3} = 2n_0^2 - n_0 + 6n_0\frac{\dot{B}_p(\text{em})\,\Omega}{B_p(\text{em})\,\dot{\Omega}} + 2\frac{\dot{B}_p(\text{em})^2\,\Omega^2}{B_p(\text{em})^2\,\dot{\Omega}^2} + 2\frac{\ddot{B}_p(\text{em})\,\Omega^2}{B_p(\text{em})\,\dot{\Omega}^2} \tag{27}$$

The quantity $m$ has been defined in terms of the observable quantities $\Omega$, $\dot{\Omega}$ and $\ddot{\Omega}$ and need not to be an integer.

When $B_p(\text{em})$ is *no function of time*, $m$ in (27) reduces to

$$m_0 = 2n_0^2 - n_0. \tag{28}$$

If $B_p(em)$ in (27) is identified with a time independent $B_p(sd)$ from (11), $n = n_0 = 3$ (see comment following (23)) and $m_0$ obtains the value

$$m_0 = 15. \tag{29}$$

If $B_p(em)$ in (27) is identified with a time dependent $B_p(sd)$ from (11), $n \ne 3$ and $n_0 = 3$ (see again comment following (23)). Substitution of $B_p(sd)$ of (11) and its time derivatives $\dot{B}_p(sd)$ and $\ddot{B}_p(sd)$ calculated from (11) into (27) yields an identity and no additional expressions for $m$ or $n$. When the observable quantities $\Omega$, $\dot{\Omega}$, $\ddot{\Omega}$ and $\dddot{\Omega}$ are known, $n$ and $m$ can be calculated from (21a) and (27a), respectively. When the value of $n$ is known the ratio $\dot{B}_p(sd)/B_p(sd)$ can be calculated from (23). Substitution of the calculated values of $m$, $n_0 = 3$ and $\dot{B}_p(sd)/B_p(sd)$ into (27) yields the ratio $\ddot{B}_p(sd)/B_p(sd)$. Since the quantity $\dddot{\Omega}$ has only been measured for PSR B1509–58 (see, [22]), only results for this pulsar will be discussed in section 6.

### 4. Magnetic field from gravito-/electromagnetic origin

Pulsars mainly consist of electrically neutral matter, probably neutrons, whereas some charged particles may also be present. In this section the charge distribution in pulsars induced by the magnetic field from gravito-magnetic origin will be investigated. Therefore, some elements of the gravito-magnetic theory will be given here. In the stationary case the magnetic induction field **B** can be calculated from the simplified gravito-magnetic equations [8, ch. 2]

$$\nabla \times \mathbf{B} = -4\pi\beta\, c^{-1}\, G^{1/2}\, \rho\, \mathbf{v} \text{ and } \nabla \cdot \mathbf{B} = 0, \tag{30}$$

where $\rho$ is the homogeneous mass density of a sphere of radius $R$ moving with a velocity $\mathbf{v} = \mathbf{\Omega} \times \mathbf{r}$ ($0 \le r \le R$). A dipolar magnetic field **B** at distance $r > R$ can be calculated from (30)

$$\mathbf{B} = \frac{3\,\mathbf{M}\cdot\mathbf{r}}{r^5}\mathbf{r} - \frac{\mathbf{M}}{r^3}, \tag{31}$$

where the gravito-magnetic dipole moment $\mathbf{M} = \mathbf{M}(gm)$ is given by (1) (see also (2)). It is assumed that $\beta = +1$, so that **M** and **S** posses opposite directions. To my knowledge the origin of the basic magnetic field **B** of pulsars has never been explained. The prediction of a such a field **B**, i. e, **B**(gm) of (31) from (30) may be considered as a first merit of the gravito-magnetic theory. Usually, it is thought that the magnetic field of pulsars is due to circulating charge (see, e. g., Reisenegger [24] for a recent discussion on the origin of the magnetic fields of pulsars). In that case, the *(electro)*magnetic field **B** in (31) can be written as **B**(em) and the magnetic dipole moment **M** as **M**(em). So far, no generally accepted model for the calculation of the field **B**(em) has, however, emerged.

The components of $\mathbf{B} = \mathbf{B}(gm)$ of (31) in spherical coordinates ($r$, $\theta$ and $\varphi$) are given by

$$\mathbf{B}_r = \frac{2M\cos\theta}{r^3}\mathbf{e}_r,\ \mathbf{B}_\theta = \frac{M\sin\theta}{r^3}\mathbf{e}_\theta \text{ and } \mathbf{B}_\varphi = 0, \quad (r > R) \tag{32}$$

where $\mathbf{e}_r$ and $\mathbf{e}_\theta$ are unit vectors. The field **B** in (31) has been calculated assuming $r > R$, but below we also need knowledge of the field **B** inside the sphere. The latter field has not yet been calculated from (30), but for a sphere with homogeneous mass density $\rho$ the magnetic field at the centre, $B_c$, can be shown to be

$$B_c = 5\,M/R^3. \tag{33}$$

The field **B** inside the sphere and its components, however, may *approximately* be calculated from (31) and (32), respectively, for all values of $0 < r \leq R$, if the magnetic dipole moment **M** of the sphere with homogeneous mass density $\rho$ shrinks to an ideal magnetic dipole moment located at the centre of the sphere. As can be seen from (33), the singularity at $r = 0$ in (31) and (32) does not occur in the full solution for $B_c$ of the equations of (30).

If **v** is the velocity of charge in the pulsar, the current density **j** is given by

$$\mathbf{j} = \sigma\,(\mathbf{E} + c^{-1}\,\mathbf{v} \times \mathbf{B}'), \text{ where } \mathbf{B}' \equiv \mathbf{B} + \mathbf{B}(\text{em}). \tag{34}$$

Here $\sigma$ is the electrical conductivity of the plasma, $\mathbf{B} = \mathbf{B}(\text{gm})$ is the magnetic induction field from gravito-magnetic origin, $\mathbf{B}(\text{em})$ the contribution from the moving charge, $\mathbf{B}' = \mathbf{B}(\text{tot})$ (see (5)) and **E** the electric field inside the sphere. Since the conductivity inside the pulsar may be extremely high, the following well-known relation from plasma physics may be used

$$\mathbf{E} = -c^{-1}\,\mathbf{v} \times \mathbf{B}' = -c^{-1}\,(\mathbf{\Omega} \times \mathbf{r}) \times \mathbf{B}'. \tag{35}$$

If accretion is present, the assumption $\mathbf{j} = 0$ underlying (35) is probably unjustified. The latter case will, however, not be considered in this work. When in a first approximation $\mathbf{B}(\text{em}) = 0$ is chosen in (34) and (35), combination of (32) and (35) yields the following components of **E** in spherical coordinates

$$\mathbf{E}_r = \frac{-\Omega\,M\,\sin^2\theta}{c\,r^2}\,\mathbf{e}_r,\quad \mathbf{E}_\theta = \frac{2\,\Omega\,M\,\sin\theta\,\cos\theta}{c\,r^2}\,\mathbf{e}_\theta \text{ and } \mathbf{E}_\varphi = 0. \tag{36}$$

Then, the charge density $\rho_e$ inside the pulsar can be calculated from the Maxwell equation

$$\nabla \cdot \mathbf{E} = 4\pi\,\rho_e. \tag{37}$$

Combination of (36) and (37) yields for the quadrupolar charge distribution $\rho_e$

$$\rho_e = \frac{\Omega\,M}{2\pi\,c\,r^3}(3\cos^2\theta - 1) = \frac{\mathbf{\Omega} \cdot \mathbf{B}}{2\pi\,c},\quad (0 \leq r \leq R) \tag{38}$$

where $M$ still equals $M(\text{gm})$. A number of authors presented results related to (38). See for a discussion, e. g., Michel and Li [25]. Note that the right hand sides of (38) do not contain the usual minus sign. The latter sign is obtained, however, if $\beta = +1$ in (1) is replaced by $\beta = -1$. The prediction of the much applied quadrupolar charge distribution $\rho_e$ of (38) may be considered as a second merit of the gravito-magnetic theory.

Integration of the quadrupolar charge distribution (38) yields a positive and a negative charge $Q_+$ and $Q_-$, respectively

$$Q_+ = \int \rho_e\,dV = \frac{3^{1/2}\,\Omega\,M}{9\,c}\ln R/R_0 = -Q_-,\quad (R_0 \leq r \leq R) \tag{39}$$

where $V$ is the volume of the sphere, $R_0$ is a chosen radius much smaller than $R$. So, for a small value of $R_0$ the net internal charge from quadrupolar origin $Q$ satisfies $Q = Q_+ + Q_- = 0$. Note that in our case the charge-separation of available plasma is caused by the *(gravito-)*magnetic field $\mathbf{B} = \mathbf{B}(\text{gm})$ of (34). Usually (see, e.g., [25]), this charge-separation is calculated from the *(electro)*magnetic field $\mathbf{B}(\text{em})$ of unspecified origin in (34).

Analogous to the derivation given by Michel and Li [25], an electric monopole in the pulsar should be adopted. Combination of $\mathbf{E}_r$ from (36) and (37) followed by inte-gration over a sphere using Gauss' law yields for the magnitude of the charge of this monopole

$$Q_{\mathrm{monopole}} = \int \rho_e \, dV = 1/(4\pi) \int (\nabla \cdot \mathbf{E}) \, dV = 1/(4\pi) \oint \mathbf{E} \cdot \mathbf{dS} = -\frac{2 \, \Omega \, M}{3 \, c}. \tag{40}$$

Contrary to Michel and Li [25], a negative charge is obtained in our calculation. A positive charge is obtained, however, when $\beta = +1$ in (1) is replaced by $\beta = -1$.

In addition, integration over the whole sphere ($0 \leq r \leq R$) shows (see, e. g. [20, § 44]) that a net magnetic dipole moment $\mathbf{M}(\mathrm{em})$ generated by the quadrupolar charge distribution (38) arises (in this case no singularity problem occurs)

$$\mathbf{M}(\mathrm{em}) = 1/(2\,c) \int \rho_e \, \mathbf{r} \times \mathbf{v} \, dV = \frac{2 \, \Omega^2 R^2}{15 \, c^2} \mathbf{M}(\mathrm{gm}). \tag{41}$$

Note that the magnetic dipole moment $\mathbf{M}(\mathrm{em})$ of (41), due to charge-separation has the same direction as the dipole moment $\mathbf{M}(\mathrm{gm})$ from gravito-magnetic origin. When $\beta = +1$ in (1) is replaced by $\beta = -1$, the signs on the right hand sides of (38) change, but the sign in (41) remains the same. So, $\mathbf{M}(\mathrm{em})$ in (41) *always enforces* $\mathbf{M}(\mathrm{gm})$. The field $\mathbf{B}(\mathrm{em})$ corresponding to $\mathbf{M}(\mathrm{em})$ of (41) is given by an expression analogous to (31).

As an aside: if the magnetic moment $\mathbf{M}$ in (31) would exclusively be due to circulating charge in the pulsar, $\mathbf{M} = \mathbf{M}(\mathrm{em})$ and $\mathbf{M}(\mathrm{gm}) = 0$. Analogously to (41), an expression for the additional *(electro)* magnetic dipole moment $\mathbf{M}'(\mathrm{em})$ can be deduced in that case

$$\mathbf{M}'(\mathrm{em}) = 1/(2\,c) \int \rho_e \, \mathbf{r} \times \mathbf{v} \, dV = \frac{2 \, \Omega^2 R^2}{15 \, c^2} \mathbf{M}(\mathrm{em}), \tag{42}$$

where again the resulting magnetic moment $\mathbf{M}'(\mathrm{em})$ *always enforces* the original $\mathbf{M}(\mathrm{em})$. It seems difficult, however, to determine the contributions $\mathbf{M}'(\mathrm{em})$ and $\mathbf{M}(\mathrm{em})$ separately.

Returning to the gravito-magnetic case, by using (41), the total magnetic moment $\mathbf{M}'$ of the pulsar can be written as

$$\mathbf{M}' = \mathbf{M}(\mathrm{gm}) + \mathbf{M}(\mathrm{em}) = (1 + \frac{2 \, \Omega^2 R^2}{15 \, c^2}) \, \mathbf{M}(\mathrm{gm}). \tag{43}$$

In chapter 4 of [8] a theoretical example is discussed in which circulating charge *weakens* a basic magnetic field from gravito-magnetic origin. In that case the resulting magnetic moments $\mathbf{M}(\mathrm{em})$ and $\mathbf{M}(\mathrm{gm})$ in (43) are anti-parallel and the total magnetic moment $\mathbf{M}'$ *is weakened* by $\mathbf{M}(\mathrm{em})$.

Analogous to the derivation of (38), by using (43), the following higher order quadrupolar charge distribution $\rho_e'$ can be deduced

$$\rho_e' = \frac{\Omega \, M'}{2\pi \, c \, r^3}(3\cos^2\theta - 1) = \frac{\mathbf{\Omega} \cdot \mathbf{B}'}{2\pi \, c}. \quad (0 \leq r \leq R) \tag{44}$$

Moreover, analogous to the derivation of (41), the following higher order magnetic dipole moment $\mathbf{M}'(\mathrm{em})$ can be deduced

$$\mathbf{M}'(\mathrm{em}) = \frac{2 \, \Omega^2 R^2}{15 \, c^2}(1 + \frac{2 \, \Omega^2 R^2}{15 \, c^2}) \, \mathbf{M}(\mathrm{gm}). \tag{45}$$

Although other mechanisms (for example: accretion, or the mechanism mentioned in the comment following (43)) may be the main cause of the *(electro)*magnetic dipole moment of (10), the magnetic dipole moment **M′**(em) of (45) may also contribute. In any case, the newly deduced magnetic dipole moment **M′**(em) of (45) may be considered as a third consequence of the gravito-magnetic theory.

Note that combination of (3a), (6b) and (45) yields an additional expression for $\beta'$

$$\beta' = + 2/15 \, (\Omega R/c)^2 \, \{1 + 2/15 \, (\Omega R/c)^2\}. \tag{46}$$

An upper limit of $\beta' = 2/15 \, (1 + 2/15) = 0.1333 \times 1.1333 = + 0.1511$ is obtained from (46), when the ratio $\Omega R/c$ approaches unity value for pulsars with a very short period. It is noticed that still higher order values for $\beta'$ can be calculated, but the series converges quickly to $+ 0.1538$. From the comment following (41) it can be concluded that the sign of $\beta'$ is positive for both $\beta = + 1$ and $\beta = - 1$.

It may be possible, that equation (44) remains valid for values of $r$ somewhat larger than $R$. In that case the quantity $R$ in (45) and (46) must be replaced by $r$. A larger $r$ results into an increasing value of $\beta'$, but the upper limit of $\beta'$ remains 0.1538 when $r$ equals $r = c/\Omega$, the classical radius of the light cylinder. A charge density according to (38) outside the pulsar radius $R$ has been considered by a number of authors. See discussions from, e.g., [21, pp. 178–181], Michel and Li [25] and Michel [26] (for recent theoretical and observational developments).

### 5. Observational data

In table 2 the values for the observed magnetic field $B_p(\text{tot})$ of a number of accretion powered X-ray, *binary* pulsars are taken from Coburn *et al.* [27] and others [28–30], whereas the corresponding values of $P$ and $\dot{P}$ are taken from the same references or other ones [31–36]. Values of $B_p(\text{tot})$ are obtained from so-called cyclotron resonance spectral features (CRSFs) that are attributed to resonant scattering of photons by electrons. Unfortunately, no generally accepted model is available for the shape of the continuum part of the X-ray spectrum, whereas such a model is necessary to isolate the line-like spectral features. Coburn *et al.* [27], however, applied the same continuum model (power-law with high-energy cutoff) to ten pulsars in table 2 and therefore their results have been used as much as possible.

More explicitly, in the presence of a gravitational field the field $B_p(\text{tot})$ has been calculated from the fundamental energy $E_{\text{cycl}}$

$$E_{\text{cycl}} = \left(\frac{e\hbar}{m_e c}\right)\left(1 - \frac{2Gm}{c^2 r}\right)^{1/2} B = 11.58 \times 10^{-12} \, 0.7657 \, B(\text{G}) \text{ keV}, \tag{47}$$

where $m_e$ is the mass of the electron, $m = 1.4 \, m_\odot$ is the mass of the pulsar and $r = 10$ km. When the scattering of the photons by the electrons takes place at the polar cap of the pulsar, the field $B(\text{G})$ in Gaussian units may be taken equal to $B_p(\text{tot})$. Since the field $B_p(\text{tot})$ can more or less directly be obtained, it is generally denoted as the observed field.

Firstly, the observed magnetic field $B_p(\text{tot})$ and the calculated field $B_p(\text{gm})$ from (4) for $\beta = 1$ in table 2 may be compared. Agreement between these fields is only approximate. If the field $B_p(\text{tot})$ would be equal to $B_p(\text{gm})$, then the quantity $\beta_{\text{eff}}$ from (6a) would yield unity value. It is noticed that larger values for $B_p(\text{gm})$ and $\beta_{\text{eff}}$ are obtained, if the angular velocity in the core, $\Omega_{\text{core}}$, of the pulsar is larger than the angular velocity at the surface, $\Omega_{\text{crust}}$, as a result of retrograde accretion. In that case $\Omega_{\text{crust}}$ in (4a) must be replaced by $\Omega_{\text{core}}$. The observed spin-up episodes in many of the pulsars in table 2 may be an indication of the validity of such an interpretation.

**Table 2. Observed and calculated magnetic fields of a number of accreting, slowly rotating, X-ray emitting, binary pulsars.**

| | Name (Type) {eclipsing} [references] | Period $P$ (s) | $\dot{P}$ (s.s$^{-1}$) | $B_p$(tot) (G) | $B_p$(gm) (G) | $B_p$(sd) (G) | $B_p$(def) (G) | $\beta_{eff}$ | $\beta'$ | Log $\tau_c$ (yr) |
|---|---|---|---|---|---|---|---|---|---|---|
| 1 | 4U 0352+309 (HMXB[a]) {no} [27, 31] | 837.7 | ~ + 4x10$^{-9}$ since 1981 (− 4x10$^{-11}$ from 1976 – 1981) | 3.2x10$^{12}$ | 6.5x10$^{10}$ | 1.2x10$^{17}$ | 3.4x10$^{10}$ | 49 | 2x10$^6$ | 3.5 |
| 2 | GX 301–2 (HMXB) {nearly} [27, 32] | 681 | Spin-up/ spin-down (irregular) | 4.8x10$^{12}$ | 8.0x10$^{10}$ | | | 60 | | |
| 3 | 4U 1538–52 (HMXB) {yes} [27, 32] | 528.8 | Spin-up/ spin-down (irregular) | 2.3x10$^{12}$ | 1.0x10$^{11}$ | | | 23 | | |
| 4 | 4U 1907+09 (HMXB) {nearly} [27, 33] | 440.6 | + 6.9x10$^{-9}$ (steady spin-down during 15 years) | 2.1x10$^{12}$ | 1.2x10$^{11}$ | 1.1x10$^{17}$ | 6.1x10$^{10}$ | 18 | 9.2x10$^5$ | 3.01 |
| 5 | Vela X-1 (HMXB) {yes} [27, 32] | 283.2 | spin-down/ spin-up (irregular) | 2.8x10$^{12}$ | 1.9x10$^{11}$ | | | 15 | | |
| 6 | A 0535+26[b] (HMXB) [28] | 103.5 | | 1.2x10$^{13}$ | 5.2x10$^{11}$ | | | 23 | | |
| 7 | XTE J1946+274 (HMXB) {no} [27] | 15.83 | | 3.9x10$^{12}$ | 3.4x10$^{12}$ | | | 1.1 | | |
| 8 | LMC X-4 (HMXB) {yes} [29] | 13.5 | + 6x10$^{-11}$ from 1985 – 1996 (short term: ~ − 10$^{-10}$) | 1.1x10$^{13}$ | 4.0x10$^{12}$ | 2x10$^{15}$ | 3x10$^{10}$ | 2.8 | 5x10$^2$ | 3.6 |
| 9 | 4U 1626–67 (LMXB[c]) {no} [27, 32, 34] | 7.67 | + 4.2x10$^{-11}$ since 1990 (− 5.0x10$^{-11}$ from 1977 – 1990) | 4.4x10$^{12}$ | 7.1x10$^{12}$ | 1.1x10$^{15}$ | 3.6x10$^{10}$ | 0.62 | 1.5x10$^2$ | 3.46 |
| 10 | Cen X-3 (HMXB) {yes} [27, 32, 35] | 4.82 | + 9x10$^{-11}$ (long term: − 2x10$^{-11}$) | 3.4x10$^{12}$ | 1.1x10$^{13}$ | 10$^{15}$ | 7x10$^{10}$ | 0.31 | 90 | 2.9 |
| 11 | V 0332+53 (HMXB) [30] | 4.37 | | 3.2x10$^{12}$ | 1.2x10$^{13}$ | | | 0.27 | | |
| 12 | 4U 0115+63 (HMXB) {no} [27] | 3.61 | | 1.3x10$^{12}$ | 1.5x10$^{13}$ | | | 0.087 | | |
| 13 | Her X-1 (LMXB) {yes} [27, 32, 36] | 1.238 | ~ + 10$^{-12}$ (long term: ~ − 10$^{-13}$) | 4.6x10$^{12}$ | 4.4x10$^{13}$ | 7x10$^{13}$ | 10$^{10}$ | 0.10 | 2 | 4.3 |

[a] HMXB = High mass X-ray binary. [b] When E$_{cycl}$ = 110 keV is attributed to the second harmonic instead of the first, $B_p$(tot) is half as large. [c] LMXB = Low mass X-ray binary.

Secondly, if no gravito-magnetic contribution to the field $B_p(tot)$ would be present at all, the spin-down magnetic field $B_p(sd)$ calculated from (11) might coincide with the field $B_p(tot)$. Table 2 shows, however, that in general the field $B_p(tot)$ is much smaller than $B_p(sd)$. Moreover, a value of $\theta < 50°$ was deduced by Bulik *et al.* [37] for a large sample of accretion powered X-ray pulsars. According to (10) and (11), values of $B_p(sd)$ in table 2 must then be considered as minima. Since spin-up episodes for most pulsars in table 2 are observed, the mechanism of magnetic dipole radiation, explicitly leading to spin-down and yielding equation (11), cannot satisfactorily explain the found values of $B_p(tot)$. In addition, the accretion process of the pulsars in table 2 may change the quantity $\dot{E}$ of (8). In that case, the derivation of $B_p(sd)$ of (11) from a combination of (8) and (9) does not apply and calculated values of $B_p(sd)$ from (11) are not reliable. The discrepancy between $B_p(tot)$ and $B_p(sd)$ may, however, also be attributed to electron scattering, not at the polar cap, but at higher altitudes above the surface of the pulsar.

When no observed value for $B_p(tot)$ for a pulsar is known, the value of $\beta_{eff}$ from (6a) cannot directly be tested (compare with table 4). One might try to deduce $\beta_{eff}$ from (7b)

$$\mathbf{B}(sd) = \mathbf{B}(em) = (\beta_{eff} - 1)\, \mathbf{B}(gm) = \beta'\, \mathbf{B}(gm), \text{ or } \beta_{eff} = 1 + \beta', \qquad (48)$$

where $B_p(gm)$ and $B_p(sd)$ may be calculated from (4) and (11), respectively. Values of $\beta'$ have been added to table 2, when data are available. Most pulsars in table 2 show irregular spin-down and spin-down regimes, so that calculation of $B_p(sd)$ and $\beta'$ will be difficult. In addition, 4U 1626–67 has the peculiarity of showing a steady spin-up from 1977 until 1990 and a steady spin-down afterwards. Only pulsar 4U 1907–09 shows a steady spin-down up to now, in agreement with the magnetic dipole radiation model. So, for many pulsars in table 2 the relation $\beta_{eff} = 1 + \beta'$ will not be valid. As pointed out above, this discrepancy may be due to electromagnetic effects (see comment following (43)) and to retrograde accretion with $\Omega_{crust} < \Omega_{core}$, or to electron-scattering at higher altitudes above the surface of the pulsar.

In addition, the field $B_p(tot)$ may be compared with the deformed magnetic field $B_p(def)$ of (17), if data are available. Agreement between $B_p(tot)$ and $B_p(def)$ in table 2 is better than between $B_p(tot)$ and $B_p(sd)$ in most cases, but worse than between $B_p(tot)$ and $B_p(gm)$. Therefore, the deformation of magnetic field lines may be relatively unimportant and a value $p = 2$ seems less likely.

Finally, the mechanism described in section 4 leading to the expression $\beta'$ of (46) seems of minor importance for the pulsars in table 2. Calculated values for $\beta'$ from (6b) in table 2 are much larger than predicted values for $\beta'$ from (46), because values of $\Omega R/c$ are small for the considered pulsars (the values for $\Omega$ are small), even in the case when $\Omega r/c$ would be equal to unity value.

In table 3 data and results are given for five *isolated* pulsars with reported values for $B_p(tot)$, analogously to table 2. Data for the soft gamma repeater SGR 1806–20 are taken from Ibrahim *et al.* [38]. Soft gamma repeaters resemble in their behaviour to the slowly rotating, X-ray emitting, binary pulsars of table 2, but, so far, no conclusive evidence for a companion or for accretion has been found for these objects. They show, however, gamma ray bursts from time to time. Since the observed absorption cyclotron features for the soft gamma repeater SGR 1806–20 are attributed to electron or proton resonance, two values for the observed magnetic field $B_p(tot)$ are presented in table 3 (see a discussion on this subject in ref. [38]). Further, data for the anomalous X-ray pulsar (= AXP) 1RXS J1708–4009 from Rea *et al.* [39] and Kaspi and Gavriil [40] are presented in table 3. AXPs are related to SGRs, but they show no gamma ray bursts up to now.

Haberl *et al.* [41] reported data for RX J0720.4–3125. Note that the value of $\dot{P}$ is rather uncertain. RX J0720.4–3125 is a member of a group X-ray dim isolated neutron stars (XDINs), which share similar properties: soft blackbody-like X-ray spectrum, radio-quiet and no known association with a super-nova remnant.

**Table 3. Observed and calculated magnetic fields of some isolated pulsars.**

| | Name (type) {origin signal} [references] | Period $P$ (s) | $\dot{P}$ (s.s$^{-1}$) | $B_p$(tot) (G) | $B_p$(gm) (G) | $B_p$(sd) (G) | $B_p$(def) (G) | $\beta_{eff}$ | $\beta'$ | Log $\tau_c$ (yr) |
|---|---|---|---|---|---|---|---|---|---|---|
| 1 | SGR 1806–20 {electron} {proton$^a$} [38] | 7.47 | 8.2x10$^{-11}$ | 5.6x10$^{11}$ 1.0x10$^{15}$ | 7.2x10$^{12}$ | 1.6x10$^{15}$ | 5.1x10$^{10}$ | 7.8x10$^{-2}$ 1.4x10$^2$ | 2.2x10$^2$ | 3.16 |
| 2 | 1RXS J1708–4009 (AXP) {electron} {proton$^a$} [39, 40] | 11.00 | 1.9x10$^{-11}$ | 9.1x10$^{11}$ 1.7x10$^{15}$ | 4.9x10$^{12}$ | 9.3x10$^{14}$ | 2.2x10$^{10}$ | 0.19 3.5x10$^2$ | 1.9x10$^2$ | 3.88 |
| 3 | RX J0720.4–3125 (XDIN) {electron} {proton$^a$} [41] | 8.39 | 5x10$^{-14}$ | 3.1x10$^{10}$ 5.6x10$^{13}$ | 6.5x10$^{12}$ | 3x10$^{13}$ | 10$^9$ | 4.8x10$^{-3}$ 8.6 | 6 | 6.4 |
| 4 | 1E 1207.4–5209 {electron} {proton$^a$} {He$^+$-line} [42, 43] | 0.424 | 1.4x10$^{-14}$ | 8.0x10$^{10}$ 1.5x10$^{14}$ 1.5x10$^{14}$ | 1.3x10$^{14}$ | 4.9x10$^{12}$ | 2.8x10$^9$ | 6.2x10$^{-4}$ 1.2 1.2 | 3.8x10$^{-2}$ | 5.7 |
| 5 | B1821–24 (INS) {electron} {proton$^a$} [44] | 0.00305 | 1.6x10$^{-18}$ | 3.7x10$^{11}$ 6.8x10$^{14}$ | 1.8x10$^{16}$ | 4.5x10$^9$ | 3.5x10$^8$ | 2.1x10$^{-5}$ 3.8x10$^{-2}$ | 2.5x10$^{-7}$ | 7.48 |

$^a$ $B_p$(tot) is calculated from: $E_{cyc1} = 6.305\text{x}10^{-15}\ 0.7657\ B(\text{G})$ keV. Compare with (47): the electron mass $m_e$ has been replaced by the proton mass $m_p$. Again, for these pulsars $m = 1.4\ m_\odot$ and $r = 10$ km have been used.

Data for the radio-quiet pulsar 1E 1207.4–5209 also showing blackbody X-radiation and probably situated in the centre of supernova remnant PKS 1209–51/52 were taken from Bignami *et al.* [42]. These authors observed three, perhaps four, distinct features in the X-ray spectrum of this star and attributed them to electron cyclotron resonance absorptions. Sanwal *et al.* [43], however, earlier observed two features in the X-ray ray spectrum and attributed them to atomic transitions of once-ionised helium. Results of suggested interpretations are given in table 3.

Becker *et al.* [44] recently reported the (marginal) detection of an electron cyclotron line in the X-ray spectrum of the isolated millisecond pulsar B1821–24. Results for this pulsar in the globular cluster M28 are given in table 3, too. As an example, the value of $\beta_{eff} = 2.1\text{x}10^{-5}$ of pulsar B1821–24 for an electron cyclotron line can be compared with $\beta_{eff} = 1$, when the field $B_p$(tot) would exclusively be due to gravito-magnetic origin. In chapter 4 of [8] a mechanism is discussed, in which circulating charge *weakens* a basic magnetic field from gravito-magnetic origin (see comment following (43)). For a series of about fourteen rotating bodies ranging from metallic cylinders in the laboratory, moons, planets, stars and the Galaxy a mean value $\bar{\beta}_{eff} = 0.076$ [8] has been found. For all these rotating bodies the discrepancy between $B_p$(tot) and $B_p$(gm) may be attributed to the cited electromagnetic weakening of the field $B_p$(gm). Moreover, one may speculate for pulsar B1821–24 that as a result of residual prograde accretion $\Omega_{core} < \Omega_{crust}$, so that $\Omega_{crust}$ in (4a) may be replaced by $\Omega_{core}$ also resulting into smaller values for $B_p$(gm) and $\beta_{eff}$.

In addition, the calculated value of $\beta' = \pm 2.5\text{x}10^{-7}$ for pulsar B1821–24 may be compared with the predicted value of $\beta' = + 6.3\text{x}10^{-4}$ from (46) for $R = 10$ km. Perhaps, the discrepancy may be explained by a change of the circulating charge (see comment following (43)). Moreover, note that the relation $\beta_{eff} = 1 + \beta'$ from (48) is not valid for this pulsar. As pointed out above, this discrepancy may again be due to electro-magnetic effects, and to prograde accretion with $\Omega_{core} < \Omega_{crust}$, or to electron-scattering at higher altitudes above the surface of the pulsar.

In view of the uncertainties in the interpretation of $E_{cycl}$, it is difficult to decide whether $B_p$(gm) or $B_p$(sd) is the best prediction for $B_p$(tot) in table 3. Further progress depends on the correct interpretation of the cyclotron features. Are they due to electron or to proton resonance?

Finally, most values of $B_p$(def) in table 3 are smaller/much smaller than those of $B_p$(tot) calculated from the electron/proton cyclotron interpretation, respectively. In the latter case a value $p = 2$ seems again less likely (compare with table 2).

In table 4 the Parkes multi-beam pulsar survey-I [45] has been chosen as a representative sample of 100 pulsars. Earlier, Woodward [19] calculated the values of $\beta'$ from $\beta' = M(sd)/M(WB) = B_p(sd)/B_p(WB)$, where WB denotes Wilson-Blackett (see formula (1)) for another sample of more than 100 pulsars and concluded that $\beta'$ is no universal constant. In order to explain this non-constancy, he suggested a secondary mechanism of magnetic field generation. Note that (24) offers an expression for $\beta' \equiv B_p(sd)/B_p(gm)$ depending on the quantities $\Omega$ and $\dot{\Omega}$, but in fact this equation is equivalent to equation (6b). In table 4 the absolute values for $\beta'$ from (6b) of the Parkes pulsar survey-I [45] are given. From these values a mean value $\overline{\beta'} = 0.12$ can be calculated.

The newly deduced *(electro)*magnetic dipole moment $M'$(em) of (45) generated by moving charge in the magnetic field of gravito-magnetic origin may partly be identified with the *(electro)*magnetic dipole moment $M$(em) of (10) following from the standard magnetic dipole radiation model. From (45) a formula for $\beta'$ given in (46) can be deduced. Note that $\beta'$ in (46) is always positive. The lower limit of $\beta'$ from (46) reduces to $\beta' = + 2/15 \, (\Omega R/c)^2$ (as an example, $\beta' = + 2.3 \times 10^{-8}$ for $P = 0.5$ s and $R = 10$ km), whereas the upper limit of $\beta'$ is given by $\beta' = + 0.15$ (see section 4). Therefore, the mean value $|\overline{\beta'}| = 0.12$ may lie in the range of values for $\beta'$ predicted by (46), although the sign of $\beta'$ is not known. However, the *(electro)*magnetic dipole moments $M'$(em) of (42) and $M$(em) itself in (42) may also contribute to $M$(em) of (10) and to $\beta'$ of (6b), so that the existence of the magnetic dipole moment **M'**(em) of (45) has not yet been proven.

Since the field $B_p$(tot) has not yet *directly* been observed for these pulsars, $B_p$(sd) of (11) might be used to estimate the value of $\beta_{eff}$ using the relation $\beta_{eff} = 1 + \beta'$ of (48). From the mean value $|\overline{\beta'}| = 0.12$ of 99 pulsars in table 4 a mean value $\overline{\beta}_{eff} = 1 + \overline{\beta'} = 1.12$ can be calculated from (48), if all values of $\beta'$ would be positive (or 0.88, if all values of $\beta'$ would be negative). According to (5) and (6), $B_p$(tot) $= B_p$(gm) implies $\beta_{eff} = 1$ and $\beta' = 0$, so that the results $\overline{\beta}_{eff} = 1.12$ or 0.88 possess an order of magnitude that may be compatible with the existence of the gravito-magnetic moment **M**(gm) of (1). Although the accretion process for the isolated pulsars of table 4 may be of minor importance, the validity of relation $\beta_{eff} = 1 + \beta'$ of (48) for these pulsars is uncertain (see comments to tables 2 and 3).

For the three short-period, binary pulsars in table 4 relatively low values for $\beta'$ are found ($\beta' \approx 10^{-5}$–$10^{-7}$). Generation of the emission at higher altitudes above the surface of the pulsar might explain the low values of $B_p$(sd) and $\beta'$ for these pulsars in table 4. According to the relation $\beta_{eff} = 1 + \beta'$ of (48), the small values of $\beta'$ would imply a value of $\beta_{eff} \approx 1$. From (6) and (48) then follows, that the *(gravito-)*magnetic field $B_p$(gm) is large compared with the *(electro)*magnetic field $B_p$(em). However, for these binary pulsars the validity of relation $\beta_{eff} = 1 + \beta'$ of (48) is also uncertain (see previous paragraph). Note that a change of the charge circulation and/or prograde accretion with $\Omega_{core} < \Omega_{crust}$ may lower the values of $B_p$(gm) and of $\beta_{eff}$ (compare with the comment to the isolated pulsar B1821–24 following table 3).

The values of $B_p$(def) in table 4 are orders of magnitude smaller than $B_p$(sd), so that again $p = 2$ seems less likely (compare with table 1).

Usually, the origin of the magnetic field of pulsars is attributed to charge moving in the pulsar, e. g., in terms of a dynamo theory. For example, Reisenegger [24] has recently given a discussion on this subject. However, no generally accepted model has emerged, so far.

Table 4. Calculated magnetic fields $B_p(gm)$ of (4) for $\beta = 1$, $B_p(sd)$ of (11) and $B_p(def)$ of (17), respectively, of the 100 pulsars of the Parkes multi-beam pulsar survey – I.

|    | PSR J     | Period $P$ (s) | $\dot{P}$ (s.s$^{-1}$) | $B_p(gm)$ (G) | $B_p(sd)$ (G) | $B_p(def)$ (G) | $\beta'$ | Log $\tau_C$ (yr) |
|----|-----------|--------|------------------|-----------|-----------|-----------|-----------|---------|
| 1  | 1307–6318 | 4.962  | 2.11x10$^{-14}$  | 1.1x10$^{13}$ | 2.1x10$^{13}$ | 1.0x10$^{9}$  | 1.9       | 6.57    |
| 2  | 1444–5941 | 2.760  | 8.2x10$^{-15}$   | 2.0x10$^{13}$ | 9.6x10$^{12}$ | 8.4x10$^{8}$  | 0.49      | 6.73    |
| 3  | 1601–5244 | 2.559  | 7.2x10$^{-16}$   | 2.1x10$^{13}$ | 2.7x10$^{12}$ | 2.6x10$^{8}$  | 0.13      | 7.75    |
| 4  | 1312–6400 | 2.437  | 6.8x10$^{-16}$   | 2.2x10$^{13}$ | 2.6x10$^{12}$ | 2.6x10$^{8}$  | 0.12      | 7.75    |
| 5  | 1303–6305 | 2.307  | 2.18x10$^{-15}$  | 2.3x10$^{13}$ | 4.5x10$^{12}$ | 4.8x10$^{8}$  | 0.20      | 7.22    |
| 6  | 1245–6238 | 2.283  | 1.09x10$^{-14}$  | 2.4x10$^{13}$ | 1.0x10$^{13}$ | 1.1x10$^{9}$  | 0.42      | 6.52    |
| 7  | 1049–5833 | 2.202  | 4.41x10$^{-15}$  | 2.5x10$^{13}$ | 6.3x10$^{12}$ | 6.9x10$^{8}$  | 0.25      | 6.90    |
| 8  | 1632–4621 | 1.709  | 7.60x10$^{-14}$  | 3.2x10$^{13}$ | 2.3x10$^{13}$ | 3.3x10$^{9}$  | 0.72      | 5.55    |
| 9  | 0838–3947 | 1.704  | 8x10$^{-16}$     | 3.2x10$^{13}$ | 2.4x10$^{12}$ | 3.4x10$^{8}$  | 7.5x10$^{-2}$ | 7.53 |
| 10 | 1001–5559 | 1.661  | 8.60x10$^{-16}$  | 3.3x10$^{13}$ | 2.4x10$^{12}$ | 3.5x10$^{8}$  | 7.3x10$^{-2}$ | 7.49 |
| 11 | 1609–5158 | 1.279  | 1.30x10$^{-14}$  | 4.2x10$^{13}$ | 8.3x10$^{12}$ | 1.6x10$^{9}$  | 0.19      | 6.19    |
| 12 | 1345–6115 | 1.253  | 3.25x10$^{-15}$  | 4.3x10$^{13}$ | 4.1x10$^{12}$ | 7.9x10$^{8}$  | 9.5x10$^{-2}$ | 6.79 |
| 13 | 1724–3505 | 1.222  | 2.11x10$^{-14}$  | 4.4x10$^{13}$ | 1.0x10$^{13}$ | 2.0x10$^{9}$  | 0.23      | 5.96    |
| 14 | 1616–5109 | 1.220  | 1.91x10$^{-14}$  | 4.4x10$^{13}$ | 9.8x10$^{12}$ | 1.9x10$^{9}$  | 0.20      | 6.01    |
| 15 | 1726–3530 | 1.110  | 1.22x10$^{-12}$  | 4.9x10$^{13}$ | 7.4x10$^{13}$ | 1.6x10$^{10}$ | 1.5       | 4.16    |
| 16 | 1621–5039 | 1.084  | 1.30x10$^{-14}$  | 5.0x10$^{13}$ | 7.6x10$^{12}$ | 1.7x10$^{9}$  | 0.15      | 6.12    |
| 17 | 1622–4944 | 1.073  | 1.71x10$^{-14}$  | 5.0x10$^{13}$ | 8.7x10$^{12}$ | 2.1x10$^{9}$  | 0.17      | 6.00    |
| 18 | 1725–3546 | 1.032  | 1.50x10$^{-14}$  | 5.2x10$^{13}$ | 8.0x10$^{12}$ | 1.9x10$^{9}$  | 0.15      | 6.04    |
| 19 | 1616–5208 | 1.026  | 2.89x10$^{-14}$  | 5.3x10$^{13}$ | 1.1x10$^{13}$ | 2.6x10$^{9}$  | 0.21      | 5.75    |
| 20 | 1605–5215 | 1.014  | 4.75x10$^{-15}$  | 5.3x10$^{13}$ | 4.4x10$^{12}$ | 1.1x10$^{9}$  | 8.3x10$^{-2}$ | 6.53 |
| 21 | 1144–6146 | 0.9878 | – 4x10$^{-17}$   | 5.5x10$^{13}$ | a | a | a | a |
| 22 | 1513–5739 | 0.9735 | 2.76x10$^{-14}$  | 5.6x10$^{13}$ | 1.0x10$^{13}$ | 2.6x10$^{9}$  | 0.19      | 5.75    |
| 23 | 1434–6029 | 0.9633 | 1.03x10$^{-15}$  | 5.6x10$^{13}$ | 2.0x10$^{12}$ | 5.1x10$^{8}$  | 3.6x10$^{-2}$ | 7.17 |
| 24 | 0922–4949 | 0.9503 | 9.76x10$^{-14}$  | 5.7x10$^{13}$ | 1.9x10$^{13}$ | 5.0x10$^{9}$  | 0.33      | 5.19    |
| 25 | 1348–6307 | 0.9278 | 3.79x10$^{-15}$  | 5.8x10$^{13}$ | 3.8x10$^{12}$ | 9.9x10$^{8}$  | 6.6x10$^{-2}$ | 6.59 |
| 26 | 1536–5433 | 0.8814 | 1.91x10$^{-15}$  | 6.1x10$^{13}$ | 2.6x10$^{12}$ | 7.2x10$^{8}$  | 4.3x10$^{-2}$ | 6.86 |
| 27 | 1649–4349 | 0.8707 | 4.4x10$^{-17}$   | 6.2x10$^{13}$ | 4.0x10$^{11}$ | 1.1x10$^{8}$  | 6.5x10$^{-3}$ | 8.50 |
| 28 | 1628–4804 | 0.8660 | 1.24x10$^{-15}$  | 6.3x10$^{13}$ | 2.1x10$^{12}$ | 5.9x10$^{8}$  | 3.3x10$^{-2}$ | 7.04 |
| 29 | 1144–6217 | 0.8507 | 3.08x10$^{-14}$  | 6.4x10$^{13}$ | 1.0x10$^{13}$ | 2.9x10$^{9}$  | 0.16      | 5.64    |
| 30 | 1252–6314 | 0.8233 | 1.1x10$^{-16}$   | 6.6x10$^{13}$ | 6.1x10$^{11}$ | 1.8x10$^{8}$  | 9.2x10$^{-3}$ | 8.07 |
| 31 | 1632–4818 | 0.8134 | 6.51x10$^{-13}$  | 6.7x10$^{13}$ | 4.7x10$^{13}$ | 1.4x10$^{10}$ | 0.70      | 4.30    |
| 32 | 1220–6318 | 0.7892 | 8x10$^{-17}$     | 6.9x10$^{13}$ | 5.1x10$^{11}$ | 1.6x10$^{8}$  | 7.4x10$^{-3}$ | 8.19 |
| 33 | 1715–3700 | 0.7796 | 1.5x10$^{-16}$   | 6.9x10$^{13}$ | 6.9x10$^{11}$ | 2.1x10$^{8}$  | 1.0x10$^{-2}$ | 7.92 |
| 34 | 1002–5559 | 0.7775 | 1.57x10$^{-15}$  | 7.0x10$^{13}$ | 2.2x10$^{12}$ | 7.0x10$^{8}$  | 3.1x10$^{-2}$ | 6.90 |
| 35 | 1429–5935 | 0.7639 | 4.28x10$^{-14}$  | 7.1x10$^{13}$ | 1.2x10$^{13}$ | 3.7x10$^{9}$  | 0.17      | 5.45    |
| 36 | 1623–4949 | 0.7257 | 4.21x10$^{-14}$  | 7.5x10$^{13}$ | 1.1x10$^{13}$ | 3.7x10$^{9}$  | 0.15      | 5.44    |
| 37 | 1407–6153 | 0.7016 | 8.85x10$^{-15}$  | 7.7x10$^{13}$ | 5.0x10$^{12}$ | 1.7x10$^{9}$  | 6.5x10$^{-2}$ | 6.10 |
| 38 | 1130–5925 | 0.6810 | 9.52x10$^{-16}$  | 8.0x10$^{13}$ | 1.6x10$^{12}$ | 5.8x10$^{8}$  | 2.0x10$^{-2}$ | 7.05 |
| 39 | 1056–5709 | 0.6761 | 5.76x10$^{-16}$  | 8.0x10$^{13}$ | 1.3x10$^{12}$ | 4.5x10$^{8}$  | 1.6x10$^{-2}$ | 7.27 |
| 40 | 1611–4949 | 0.6664 | 5.4x10$^{-16}$   | 8.1x10$^{13}$ | 1.2x10$^{12}$ | 4.4x10$^{8}$  | 1.5x10$^{-2}$ | 7.29 |
| 41 | 1613–5234 | 0.6552 | 6.63x10$^{-15}$  | 8.3x10$^{13}$ | 4.2x10$^{12}$ | 1.6x10$^{9}$  | 5.1x10$^{-2}$ | 6.19 |
| 42 | 1123–6102 | 0.6402 | 6.46x10$^{-15}$  | 8.5x10$^{13}$ | 4.1x10$^{12}$ | 1.6x10$^{9}$  | 4.9x10$^{-2}$ | 6.20 |
| 43 | 1716–3720 | 0.6303 | 1.80x10$^{-14}$  | 8.6x10$^{13}$ | 6.8x10$^{12}$ | 2.6x10$^{9}$  | 7.9x10$^{-2}$ | 5.74 |
| 44 | 1341–6023 | 0.6273 | 1.95x10$^{-14}$  | 8.6x10$^{13}$ | 7.1x10$^{12}$ | 2.7x10$^{9}$  | 8.3x10$^{-2}$ | 5.71 |
| 45 | 1309–6415 | 0.6195 | 8.79x10$^{-15}$  | 8.7x10$^{13}$ | 4.7x10$^{12}$ | 1.8x10$^{9}$  | 5.4x10$^{-2}$ | 6.05 |
| 46 | 1728–3733 | 0.6155 | 7x10$^{-17}$     | 8.8x10$^{13}$ | 4.2x10$^{11}$ | 1.7x10$^{8}$  | 4.8x10$^{-2}$ | 8.14 |
| 47 | 1540–5736 | 0.6129 | 4.2x10$^{-16}$   | 8.8x10$^{13}$ | 1.0x10$^{12}$ | 4.1x10$^{8}$  | 1.1x10$^{-2}$ | 7.36 |
| 48 | 1653–4249 | 0.6126 | 4.81x10$^{-15}$  | 8.8x10$^{13}$ | 3.5x10$^{12}$ | 1.4x10$^{9}$  | 4.0x10$^{-2}$ | 6.30 |
| 49 | 1347–5947 | 0.6100 | 1.42x10$^{-14}$  | 8.9x10$^{13}$ | 6.0x10$^{12}$ | 2.4x10$^{9}$  | 6.7x10$^{-2}$ | 5.83 |
| 50 | 1558–5419 | 0.5946 | 6.04x10$^{-15}$  | 9.1x10$^{13}$ | 3.8x10$^{12}$ | 1.6x10$^{9}$  | 4.2x10$^{-2}$ | 6.19 |

a Since $\dot{P}$ is negative (the pulsar spins up), equation (11) is not applicable.

**Table 4** *(Cont.)*. Calculated magnetic fields $B_p$(gm) of (4) for $\beta = 1$, $B_p$(sd) of (11) and $B_p$(def) of (17), respectively, of the 100 pulsars of the Parkes multi-beam pulsar survey – I.

|  | PSR J | Period $P$ (s) | $\dot{P}$ (s.s$^{-1}$) | $B_p$(gm) (G) | $B_p$(sd) (G) | $B_p$(def) (G) | $\beta'$ | Log $\tau_C$ (yr) |
|---|---|---|---|---|---|---|---|---|
| 51 | 1709–3841 | 0.5870 | 7.86x10$^{-15}$ | 9.2x10$^{13}$ | 4.3x10$^{12}$ | 1.8x10$^{9}$ | 4.7x10$^{-2}$ | 6.07 |
| 52 | 1224–6208 | 0.5858 | 2.02x10$^{-14}$ | 9.2x10$^{13}$ | 7.0x10$^{12}$ | 2.9x10$^{9}$ | 7.5x10$^{-2}$ | 5.66 |
| 53 | 1142–6230 | 0.5584 | 8x10$^{-17}$ | 9.7x10$^{13}$ | 4.3x10$^{11}$ | 1.9x10$^{8}$ | 4.4x10$^{-3}$ | 8.04 |
| 54 | 0835–3707 | 0.5414 | 9.78x10$^{-15}$ | 1.0x10$^{14}$ | 4.7x10$^{12}$ | 2.1x10$^{9}$ | 4.7x10$^{-2}$ | 5.94 |
| 55 | 1412–6111 | 0.5292 | 1.91x10$^{-15}$ | 1.0x10$^{14}$ | 2.0x10$^{12}$ | 9.3x10$^{8}$ | 2.0x10$^{-2}$ | 6.64 |
| 56 | 1322–6241 | 0.5061 | 2.59x10$^{-15}$ | 1.1x10$^{14}$ | 2.3x10$^{12}$ | 1.1x10$^{9}$ | 2.1x10$^{-2}$ | 6.49 |
| 57 | 1425–6210 | 0.5017 | 4.8x10$^{-16}$ | 1.1x10$^{14}$ | 9.9x10$^{11}$ | 4.8x10$^{8}$ | 9.0x10$^{-3}$ | 7.22 |
| 58 | 1407–6048 | 0.4923 | 3.16x10$^{-15}$ | 1.1x10$^{14}$ | 2.5x10$^{12}$ | 1.2x10$^{9}$ | 2.3x10$^{-2}$ | 6.39 |
| 59 | 1610–5006 | 0.4811 | 1.36x10$^{-14}$ | 1.1x10$^{14}$ | 5.2x10$^{12}$ | 2.6x10$^{9}$ | 4.7x10$^{-2}$ | 5.75 |
| 60 | 1305–6256 | 0.4782 | 2.11x10$^{-15}$ | 1.1x10$^{14}$ | 2.0x10$^{12}$ | 1.0x10$^{9}$ | 1.8x10$^{-2}$ | 6.56 |
| 61 | 0954–5430 | 0.4728 | 4.39x10$^{-14}$ | 1.1x10$^{14}$ | 9.2x10$^{12}$ | 4.7x10$^{9}$ | 8.1x10$^{-2}$ | 5.23 |
| 62 | 1625–4904 | 0.4603 | 1.68x10$^{-14}$ | 1.2x10$^{14}$ | 5.6x10$^{12}$ | 3.0x10$^{9}$ | 4.7x10$^{-2}$ | 5.64 |
| 63 | 1613–5211 | 0.4575 | 1.92x10$^{-14}$ | 1.2x10$^{14}$ | 6.0x10$^{12}$ | 3.2x10$^{9}$ | 5.0x10$^{-2}$ | 5.58 |
| 64 | 0901–4624 | 0.4420 | 8.75x10$^{-14}$ | 1.2x10$^{14}$ | 1.3x10$^{13}$ | 6.9x10$^{9}$ | 0.11 | 4.90 |
| 65 | 1537–5645 | 0.4305 | 2.78x10$^{-15}$ | 1.3x10$^{14}$ | 2.2x10$^{12}$ | 1.2x10$^{9}$ | 1.7x10$^{-2}$ | 6.39 |
| 66 | 1305–6203 | 0.4278 | 3.21x10$^{-14}$ | 1.3x10$^{14}$ | 7.5x10$^{12}$ | 4.2x10$^{9}$ | 5.8x10$^{-2}$ | 5.32 |
| 67 | 1119–6127 | 0.4077 | 4.02x10$^{-12}$ | 1.3x10$^{14}$ | 8.2x10$^{13}$ | 4.9x10$^{10}$ | 0.63 | 3.21 |
| 68 | 1452–5851 | 0.3866 | 5.07x10$^{-14}$ | 1.4x10$^{14}$ | 9.0x10$^{12}$ | 5.6x10$^{9}$ | 6.4x10$^{-2}$ | 5.08 |
| 69 | 1650–4502 | 0.3809 | 1.61x10$^{-14}$ | 1.4x10$^{14}$ | 5.0x10$^{12}$ | 3.2x10$^{9}$ | 3.6x10$^{-2}$ | 5.57 |
| 70 | 1543–5459 | 0.3771 | 5.20x10$^{-14}$ | 1.4x10$^{14}$ | 9.0x10$^{12}$ | 5.8x10$^{9}$ | 6.4x10$^{-2}$ | 5.06 |
| 71 | 1216–6223 | 0.3740 | 1.68x10$^{-14}$ | 1.4x10$^{14}$ | 5.1x10$^{12}$ | 3.3x10$^{9}$ | 3.5x10$^{-2}$ | 5.55 |
| 72 | 1720–3659 | 0.3511 | 3.27x10$^{-17}$ | 1.6x10$^{14}$ | 2.2x10$^{11}$ | 1.5x10$^{8}$ | 1.4x10$^{-3}$ | 8.23 |
| 73 | 1607–5140 | 0.3427 | 2.54x10$^{-15}$ | 1.7x10$^{14}$ | 1.9x10$^{12}$ | 1.3x10$^{9}$ | 1.1x10$^{-2}$ | 6.33 |
| 74 | 1412–6145 | 0.3152 | 9.87x10$^{-14}$ | 1.7x10$^{14}$ | 1.1x10$^{13}$ | 8.7x10$^{9}$ | 6.5x10$^{-2}$ | 4.70 |
| 75 | 1649–4729 | 0.2977 | 6.55x10$^{-15}$ | 1.8x10$^{14}$ | 2.8x10$^{12}$ | 2.3x10$^{9}$ | 1.6x10$^{-2}$ | 5.86 |
| 76 | 1416–6037 | 0.2956 | 4.28x10$^{-15}$ | 1.8x10$^{14}$ | 2.3x10$^{12}$ | 1.9x10$^{9}$ | 1.3x10$^{-2}$ | 6.04 |
| 77 | 1626–4807 | 0.2939 | 1.75x10$^{-14}$ | 1.8x10$^{14}$ | 4.6x10$^{12}$ | 3.8x10$^{9}$ | 2.6x10$^{-2}$ | 5.43 |
| 78 | 1413–6222 | 0.2924 | 2.23x10$^{-15}$ | 1.9x10$^{14}$ | 1.6x10$^{12}$ | 1.4x10$^{9}$ | 8.4x10$^{-3}$ | 6.32 |
| 79 | 1601–5335 | 0.2885 | 6.24x10$^{-14}$ | 1.9x10$^{14}$ | 8.6x10$^{12}$ | 7.2x10$^{9}$ | 4.5x10$^{-2}$ | 4.86 |
| 80 | 1726–3635 | 0.2874 | 1.44x10$^{-15}$ | 1.9x10$^{14}$ | 1.3x10$^{12}$ | 1.1x10$^{9}$ | 6.8x10$^{-3}$ | 6.50 |
| 81 | 1327–6400 | 0.2807 | 3.12x10$^{-14}$ | 1.9x10$^{14}$ | 6.0x10$^{12}$ | 5.2x10$^{9}$ | 3.1x10$^{-2}$ | 5.15 |
| 82 | 1530–5327 | 0.2790 | 4.68x10$^{-15}$ | 1.9x10$^{14}$ | 2.3x10$^{12}$ | 2.0x10$^{9}$ | 1.2x10$^{-2}$ | 5.98 |
| 83 | 1538–5438 | 0.2767 | 1.42x10$^{-15}$ | 2.0x10$^{14}$ | 1.3x10$^{12}$ | 1.1x10$^{9}$ | 6.5x10$^{-3}$ | 6.49 |
| 84 | 1622–4802 | 0.2651 | 3.07x10$^{-16}$ | 2.0x10$^{14}$ | 5.8x10$^{11}$ | 5.3x10$^{8}$ | 2.9x10$^{-3}$ | 7.14 |
| 85 | 1317–6302 | 0.2613 | 1.02x10$^{-16}$ | 2.1x10$^{14}$ | 3.3x10$^{11}$ | 3.1x10$^{8}$ | 1.6x10$^{-3}$ | 7.61 |
| 86 | 1115–6052 | 0.2598 | 7.24x10$^{-15}$ | 2.1x10$^{14}$ | 2.8x10$^{12}$ | 2.6x10$^{9}$ | 1.3x10$^{-2}$ | 5.76 |
| 87 | 1349–6130 | 0.2594 | 5.13x10$^{-15}$ | 2.1x10$^{14}$ | 2.3x10$^{12}$ | 2.2x10$^{9}$ | 1.1x10$^{-2}$ | 5.90 |
| 88 | 1406–6121 | 0.2131 | 5.47x10$^{-14}$ | 2.5x10$^{14}$ | 6.9x10$^{12}$ | 7.8x10$^{9}$ | 2.8x10$^{-2}$ | 4.79 |
| 89 | 0957–5432 | 0.2036 | 1.95x10$^{-15}$ | 2.7x10$^{14}$ | 1.3x10$^{12}$ | 1.5x10$^{9}$ | 4.8x10$^{-3}$ | 6.22 |
| 90 | 1723–3659 | 0.2027 | 8.01x10$^{-15}$ | 2.7x10$^{14}$ | 2.6x10$^{12}$ | 3.1x10$^{9}$ | 9.6x10$^{-3}$ | 5.60 |
| 91 | 1301–6305 | 0.1845 | 2.67x10$^{-13}$ | 2.9x10$^{14}$ | 1.4x10$^{13}$ | 1.9x10$^{10}$ | 4.8x10$^{-2}$ | 4.04 |
| 92 | 1548–5607 | 0.1709 | 1.07x10$^{-14}$ | 3.2x10$^{14}$ | 2.7x10$^{12}$ | 3.9x10$^{9}$ | 8.6x10$^{-3}$ | 5.40 |
| 93 | 1138–6207 | 0.1176 | 1.25x10$^{-14}$ | 4.6x10$^{14}$ | 2.5x10$^{12}$ | 5.0x10$^{9}$ | 5.3x10$^{-3}$ | 5.17 |
| 94 | 1232–6501$^a$ | 8.828x10$^{-2}$ | 8.1x10$^{-19}$ | 6.1x10$^{14}$ | 1.7x10$^{10}$ | 4.7x10$^{7}$ | 2.8x10$^{-5}$ | 9.24 |
| 95 | 1016–5819 | 8.783x10$^{-2}$ | 6.98x10$^{-16}$ | 6.2x10$^{14}$ | 5.0x10$^{11}$ | 1.4x10$^{9}$ | 8.1x10$^{-4}$ | 6.30 |
| 96 | 0940–5428 | 8.755x10$^{-2}$ | 3.29x10$^{-14}$ | 6.2x10$^{14}$ | 3.4x10$^{12}$ | 9.5x10$^{9}$ | 5.6x10$^{-3}$ | 4.62 |
| 97 | 1718–3825 | 7.467x10$^{-2}$ | 1.32x10$^{-14}$ | 7.3x10$^{14}$ | 2.0x10$^{12}$ | 6.5x10$^{9}$ | 2.8x10$^{-3}$ | 4.95 |
| 98 | 1112–6103 | 6.496x10$^{-2}$ | 3.15x10$^{-14}$ | 8.3x10$^{14}$ | 2.9x10$^{12}$ | 1.1x10$^{10}$ | 3.5x10$^{-3}$ | 4.51 |
| 99 | 1454–5846$^a$ | 4.525x10$^{-2}$ | 8.16x10$^{-19}$ | 1.2x10$^{15}$ | 1.2x10$^{10}$ | 6.6x10$^{7}$ | 1.0x10$^{-5}$ | 8.94 |
| 100 | 1435–6100$^a$ | 9.348x10$^{-3}$ | 2.45x10$^{-20}$ | 5.8x10$^{15}$ | 9.7x10$^{8}$ | 2.5x10$^{7}$ | 1.7x10$^{-7}$ | 9.78 |

$^a$ Binary pulsar.

# 6. Braking indices for some young pulsars

Data from several authors [21, 22, 46–48] of five isolated, young pulsars have been gathered in table 5. The values of the braking indices have all been calculated from $n = \Omega \ddot{\Omega}/\dot{\Omega}^2$. As can be seen from (21) and (22), the braking index $n$ does not directly depend on the gravito-magnetic field $B_p(gm)$ or on $\dot{B}_p(gm)$. Therefore, the validity of the gravito-magnetic hypothesis cannot been tested by considering $n$.

**Table 5. Braking index and other data of some isolated, young pulsars.**

| Name PSR [references] | Period $P$ (s) | $\dot{P}$ (s.s$^{-1}$) | $n$ | $B_p(gm)$ (G) | $B_p(sd)$ (G) | $B_p(def)$ (G) | $\beta'$ | $\tau_c$ (kyr) |
|---|---|---|---|---|---|---|---|---|
| J1119–6127 [46] | 0.4076 | 4.02x10$^{-12}$ | 2.91 ± 0.05 | 1.3x10$^{14}$ | 8.2x10$^{13}$ | 4.9x10$^{10}$ | 0.63 | 1.61 |
| B1509–58 [22] | 0.1507 | 1.54x10$^{-12}$ | 2.837 ± 0.001 | 3.6x10$^{14}$ | 3.1x10$^{13}$ | 5.0x10$^{10}$ | 0.086 | 1.55 |
| B0833–45 Vela pulsar [47] | 0.0893 | 1.25x10$^{-13}$ | 1.4 ± 0.2 | 6.1x10$^{14}$ | 6.8x10$^{12}$ | 1.8x10$^{10}$ | 0.011 | 11.3 |
| B0540–69 [48] | 0.05046 | 4.79x10$^{-13}$ | 2.125 ± 0.001 | 1.1x10$^{15}$ | 9.9x10$^{12}$ | 4.8x10$^{10}$ | 0.0090 | 1.67 |
| B0531+21 Crab pulsar [21] | 0.0331 | 4.23x10$^{-13}$ | 2.515 ± 0.005 | 1.6x10$^{15}$ | 7.6x10$^{12}$ | 5.5x10$^{10}$ | 0.0048 | 1.24 |

From observations of PSR B1509–58 carried out by Kaspi *et al.* [22] a value $n = 2.837$ was found for the first order braking index and a value $m = 14.5 \pm 3.6$ for the second order braking index. When the magnetic field $B_p(em)$ is no function of time, these empirical results may be compared with the value $n_0 = 3$ (see comment following (23)) and $m_0 = 15$ from (29), respectively. Alternatively, if it is assumed that $B_p(em) = B_p(sd)$ is time dependent, substitution of $n = 2.837$ into (23) yields a value of $\dot{B}_p(sd)/B_p(sd) = + 1/(38.0$ kyr$)$. Substitution of this ratio of $\dot{B}_p(sd)/B_p(sd)$ into (27), together with $m = 14.5$ and $n_0 = 3$, then yields a value of $\ddot{B}_p(sd)/B_p(sd) = + 1/(4.49$ kyr$)^2$.

Instead of *differentiation*, Johnston and Galloway [23] used *integration* to derive an alternative braking index $n_0$ from $\dot{\Omega} = - K \Omega^{n_0} = - k B_p(em)^2 \Omega^{n_0}$ (see (20): $K$, $n_0$, $k$ and $B_p(em)$ are assumed not to depend on time). From their formula they found $n_0 = 2.502$ for the Crab pulsar compared with $n = 2.509$ directly calculated from $n = \Omega \ddot{\Omega}/\dot{\Omega}^2$; so $n - n_0 = + 0.007$. Likewise, for PSR B1509–58 they found $n_0 = 2.80$ compared with $n = 2.837$; so $n - n_0 = + 0.037$. Thus, $n$ and $n_0$ nearly coincide for both pulsars, when $K$ does not depend on time. However, if $B_p(em) = B_p(sd)$ from (11) is time dependent, $K$ is time dependent and combination of (22) and (23) yields $n_0 = 3$. Then the values of $n$ and $n_0$ become different and a ratio $\dot{B}_p(sd)/B_p(sd)$ different from zero can be calculated from (23). Taking $n = 2.509$ for the Crab pulsar, a value of $\dot{B}_p(sd)/B_p(sd) = + 1/(10.1$ kyr$)$ can be calculated. Likewise, for PSR B1509–58 a value $\dot{B}_p(sd)/B_p(sd) = + 1/(38.0$ kyr$)$ already been found above. Thus, the absolute value of $B_p(sd)$ of both young pulsars increases in time.

Since the true age of the Crab pulsar is known (920 years in 1974 from [21]) and the characteristic time $\tau_c = 1240$ years, $\Omega_i$ can be calculated from (26). One finds $\Omega_i = 1.7\ \Omega$, or $P_i = 19$ ms. Note that the value $n = 2.515$ has been introduced into (26), instead of the canonical value $n_0 = 3$. In the latter case, $\Omega_i = 1.9\ \Omega$, or $P_i = 22$ ms (compare with [21, pp. 111–112])].

# 7. Conclusions

Following Woodward [19], magnetic fields of pulsars are compared with the prediction of the so-called Wilson-Blackett formula (1). This expression suggests a gravitational origin of the magnetic field of rotating, celestial bodies. Several attempts have been made to derive (1) in this form or another from a more general theory [8–14]. Equation (1) may be considered as a consequence of general relativity, e. g., in the gravito-magnetic approach [8]. The gravito-magnetic prediction for the magnetic induction field at the pole of the pulsar, $B_p(gm)$, is given by (4). In this work the latter field has been compared with the observed field, $B_p(tot)$, when available, or otherwise with the value of $B_p(tot)$ deduced from the magnetic field, $B_p(sd)$ of (11), predicted by the standard spin-down radiation model.

The validity of (1) is tested in table 2 by comparison of the field $B_p(gm)$ from (4) with the corresponding field $B_p(tot)$ of thirteen young, X-ray emitting, *binary* pulsars. For these pulsars the ratio $\beta_{eff} \equiv B_p(tot)/B_p(gm)$ varies from 0.1 to 60. The agreement is only approximate but much better than between $B_p(tot)$ and $B_p(sd)$ of (11).

In table 3 values of $B_p(gm)$ of (4) and $B_p(sd)$ of (11) are compared with the corresponding values of $B_p(tot)$ of five *isolated* pulsars (one soft gamma repeater, one anomalous X-ray pulsar and three isolated neutron stars). As an example, the value of $\beta_{eff} = 2.1 \times 10^{-5}$ of the *millisecond* pulsar B1821–24 for an electron cyclotron line can be compared with $\beta_{eff} = 1$, when the field $B_p(tot)$ would exclusively be due to gravito-magnetic origin. In chapter 4 of [8] a mechanism is discussed, in which circulating charge *weakens* a basic magnetic field from gravito-magnetic origin. For a series of about fourteen rotating bodies ranging from metallic cylinders in the laboratory, moons, planets, stars and the Galaxy a mean value $\bar{\beta}_{eff} = 0.076$ [8] has been found. For all these rotating bodies the discrepancy between $B_p(tot)$ and $B_p(gm)$ may be attributed to this electro-magnetic weakening of the field $B_p(gm)$. Finally, in view of the uncertainties in the identification of the measured values of $B_p(tot)$, it is difficult to decide whether $B_p(gm)$ or $B_p(sd)$ is the best prediction in table 3.

In table 4 data of a sample of 100 pulsars of the Parkes multi-beam survey - I [45] have been given. Since no values of $B_p(tot)$ are available, $B_p(sd)$ of (11) might be used to estimate the strength of the magnetic field of these pulsars from the relation $\beta_{eff} = 1 + \beta'$ of (48). A mean value $|\bar{\beta}'| = 0.12$ from $\beta' \equiv B_p(sd)/B_p(gm)$ of (6b) and the limits $\bar{\beta}_{eff} = 1 + \bar{\beta}' = 1.12$ and $0.88$ can be calculated from the 99 pulsars in table 4. So, the mean value $\bar{\beta}_{eff}$ and the gravito-magnetic prediction $\beta_{eff} = 1$ are in reasonable agreement. The results from tables 2 and 3 show, however, that the relation $\beta_{eff} = 1 + \beta'$ of (48) need not to be valid for several reasons mentioned.

More consequences of the gravito-magnetic theory for pulsars are considered in section 4. From this theory the standard quadrupolar charge density for pulsars can be derived. Moreover, the new *(electro)*magnetic dipole moment $M'(em)$ of (45), generated by circulating charge in the magnetic field of gravito-magnetic origin has been deduced. The *(electro)*magnetic dipole moment $M(em)$ of (10), following from the standard magnetic dipole radiation model, may partly be identified with this new magnetic dipole moment $M'(em)$. From (45) a formula for $\beta'$ given in (46) can be deduced. Note that $\beta'$ in (46) is always positive. Calculated values of $\beta'$ lie in between zero value and $\beta' = +0.15$ (see section 4), reasonably close to the mean value $|\bar{\beta}'| = 0.12$. More evidence is necessary, however, to prove the existence of the magnetic dipole moment $\mathbf{M}'(em)$ of (45).

Special attention has been given on the influence of the magnetic field on first order braking index $n$ (equations (21) and (22)) and second order braking index $m$ (equations (27)–(29)). In deriving these equations, it has been assumed that the moment of inertia $I$, radius $r$ and mass $m$ of the pulsar and angle $\theta$ between magnetic moment $\mathbf{M}$ and angular momentum $\mathbf{S}$ of the pulsar do not depend on time. Of course, these assumptions need not to be true. Observed values of $n$ of five young pulsars and one value for $m$ of PSR B1509–58 have been compared with theoretical predictions. Since *n* does not directly depend on the gravito-magnetic field $B_p(gm)$, *the validity of the gravito-magnetic hypothesis cannot directly be tested by considering n.*

In conclusion, one may say that the so-called Wilson-Blackett formula (1) predicts the correct order of magnitude of the magnetic field of many pulsars. Several effects from electromagnetic origin may account for discrepancies. Some new consequences from the gravito-magnetic theory for pulsars were deduced. So far, the proposed gravito-magnetic hypothesis may be compatible with known observational and theoretical evidence. As usual, many questions remain and new ones arise. A main advantage of the presented gravito-magnetic theory is, however, that it explains the origin of the magnetic fields of pulsars, and in fact of all rotating, electrically neutral bodies. Finally, conformation of the presented gravito-magnetic hypothesis would be a step forward towards further unification of existing physical theories [49].